\documentclass[10pt,conference]{IEEEtran}

\usepackage{amsmath}
\usepackage{amssymb}
\usepackage{color}

\newtheorem{lemma}{Lemma}
\newtheorem{example}{Example}
\newtheorem{remark}{Remark}
\newtheorem{definition}{Definition}
\newtheorem{theorem}{Theorem}

\newtheorem{corollary}{Corollary}
\newcommand{\C}{\mathbf{C}}
\newcommand{\F}[0]{{\mathbb{F}}}
\newcommand{\N}{\mathbf{N}}

\newcommand{\onemat}[0]{{\mathbf 1}}

\newcommand{\Hom}{\textup{Hom}}
\newcommand{\Harm}{\textup{Harm}}

\def\ket#1{\left|#1\right>}
\def\bra#1{\left<#1\right|}

\newcommand{\nix}[1]{{}}

\def\tr{\mathop{{\rm tr}}\nolimits}

\newcommand{\scal}[2]{\langle #1|#2\rangle}

\begin{document}

\title{Mutually Unbiased Bases are Complex Projective $2$-Designs}

\author{\authorblockN{Andreas Klappenecker}
\authorblockA{Department of Computer Science\\
Texas A\&M University\\
College Station, TX, 77843--3112, USA\\
Email: klappi@cs.tamu.edu}
\and
\authorblockN{Martin R{\"o}tteler}
\authorblockA{NEC Labs America, Inc.\\
4 Independence Way\\
Princeton, NJ 08540 U.S.A.\\
Email: mroetteler@nec-labs.com}
}

\maketitle

\begin{abstract}
Mutually unbiased bases (MUBs) are a primitive used in quantum
information processing to capture the principle of
complementarity. While constructions of maximal sets of $d+1$ such
bases are known for system of prime power dimension $d$, it is unknown
whether this bound can be achieved for any non-prime power dimension.
In this paper we demonstrate that maximal sets of MUBs come with a
rich combinatorial structure by showing that they actually are the
same objects as the complex projective $2$-designs with angle set
$\{0,1/d\}$. We also give a new and simple proof that symmetric
informationally complete POVMs are complex projective $2$-designs
with angle set $\{1/{(d{+}1)}\}$.
\end{abstract}

\section{Introduction}

Two quantum mechanical observables are called complementary if and
only if precise knowledge of one of them implies that all possible
outcomes are equally probable when measuring the other, see for
example~\cite[p.~561]{scully97}. The principle of complementarity was
introduced by Bohr~\cite{bohr28} in 1928, and it had a profound impact
on the further development of quantum mechanics.  A recent
application is the quantum key exchange protocol by Bennett and
Brassard~\cite{bennett84} that exploits complementarity to secure the
key exchange against eavesdropping.

We mention a simple mathematical consequence of this complementarity
principle, which motivates some key notion.  Suppose that $O$ and
$O'$ are two hermitian $d\times d$ matrices representing a pair of
complementary observables.  We assume that the eigenvalues of both
matrices are multiplicity free. It follows that the observables $O$
and $O'$ respectively have orthonormal eigenbases $B$ and $B'$ with
basis vectors uniquely determined up to a scalar factor. 

The complementarity of $O$ and $O'$ implies that if a quantum system
is prepared in an eigenstate $b'$ of the observable $O'$, and $O$ is
subsequently measured, then the probability to find the system after
the measurement in the state $b\in B$ is given by
$|\scal{b}{b'}|^2=1/d$.  Recall that two orthonormal bases $B$ and
$B'$ of~$\C^d$ are said to be \textsl{mutually unbiased} precisely
when $|\scal{b}{b'}|^2=1/d$ holds for all $b\in B$ and $b'\in
B'$. Thus the eigenbases of non-degenerate complementary observables
are mutually unbiased. Conversely, we can associate to a pair of
mutually unbiased bases a pair of non-degenerate complementary
observables.

There is a fundamental property of mutually unbiased bases that is
invaluable in quantum information processing. Suppose that we want to
determine the density matrix $\rho$ of an ensemble of quantum systems
using as few non-degenerate observables as possible. We assume that it
is possible to make a complete measurement of each observable
$O=\sum_{b\in B} x_b \ket{b}\!\bra{b\,}$, meaning that the statistics
$\tr(\rho \ket{b}\!\bra{b\,})=\bra{b}\rho\ket{b}$ is known for each
eigenvalue $x_b$ in the spectral decomposition. Ivanovi{\'c} showed in
\cite{Ivanovic:81} that complete measurements of at least $d+1$
observables are needed to reconstruct the density matrix.  He also
showed that this lower bound is attained when $d+1$ non-degenerate
pairwise complementary observables are used.

A simple example is provided by the Pauli spin matrices $\sigma_x$,
$\sigma_y$, $\sigma_z$. A complete measurement of these three
observables allows to reconstruct a $2\times 2$ density matrix, a fact
apparently known to Schwinger~\cite{schwinger60}. Nowadays, we know
how to do this state tomography process---at least in principle---in
dimensions $d=3, 4,$ and 5. It is an open problem whether it is
possible to perform this kind of state tomography in dimension~$6$,
because the construction of a set of 7 mutually unbiased bases in
dimension $d=6$ is elusive.

\section{Mutually Unbiased Bases}

\begin{definition} Two orthonormal bases $B$ and $C$ of $\C^d$ are
called mutually unbiased iff $|\scal{ b }{ c }|^2
=1/d$ holds for all $b\in B$ and $c\in C$.
\end{definition}

The goal is to construct $d+1$ mutually unbiased bases (MUBs) in any
dimension $d\geq 2$. There are several constructions known to obtain
MUBs. At least for prime power dimension the problem is completely
solved. This follows from Constructions I-III below. However, in any
dimension other than a prime power it is unknown if a maximal set of
$d+1$ MUBs can be found. The best known result is Construction IV
below which only works in dimensions $d$ which are squares and never
gives a maximal set of MUBs.
\smallskip

\textit{Construction I} (Wootters and Fields \cite{WF:89})
Let $q$ be an odd prime power. Define
\[
 \ket{v_{a,b}} = q^{-1/2} (\omega_p^{\tr(ax^2+bx)})_{x\in
\F_q}\in \C^q,
\] 
with $\omega_p=\exp(2\pi i/p)$.  Then the standard basis together with
the bases $B_a=\{ \ket{v_{a,b}}| b\in \F_q\}$, $a\in \F_q$, form a set
of $q+1$ mutually unbiased bases of $\C^{q}$.
\smallskip 

{\em Construction II} (Galois Rings \cite{KR:2004}) Let
$\textup{GR}(4,n)$ be a finite Galois ring with Teichm\"uller set
${\cal T}_n$. Define
\[
 \ket{v_{a,b}}=2^{-n/2} \left(\exp\left(\frac{2\pi i}{4}
{\tr(a+2b)x}\right)\right)_{ x\in {\cal T}_n}.
\]
Then the standard basis together with the bases 
$M_a=\{\ket{v_{a,b}} | b\in {\cal T}_n\}$,
$a\in {\cal T}_n$, form a set of $2^n+1$ mutually unbiased bases of
$\C^{2^n}$.
\smallskip

{\em Construction III} (Bandyopadhyay et al. \cite{BBRV:2001})
Suppose there exist subsets ${\cal C}_1, \ldots, {\cal C}_m$ of a
unitary error basis ${\cal B}$ such that $|{\cal C}_i| = d$, ${\cal
C}_i \cap {\cal C}_j = \{ \mathbf{1}_d \}$ for $i\not=j$, and the
elements of ${\cal C}_i$ pairwise commute. Let $M_i$ be a matrix which
diagonalizes ${\cal C}_i$. Then $M_1, \ldots, M_m$ are MUBs.
\smallskip

{\em Construction IV} (Wocjan and Beth \cite{WB:2004}) Suppose there
are $w$ mutually orthogonal Latin squares \cite{BJL1:99}, each of size
$d\times d$ over the symbol set $S=\{1,\ldots,d\}$. Then $w+2$ MUBs in
dimension $d^2$ can be constructed as follows. With each Latin square
$L$ (and additionally the square $(1,2,\ldots,n)^t \otimes
(1,\ldots,1)$) we can associate vectors of length $d$ over the
alphabet $\{1,\ldots,d^2\}$: for each symbol $\alpha \in S$ define a
vector $s_{L,\alpha} \in \C^{d}$ as follows: start with the empty list
$s_{L,\alpha}=\emptyset$. Then traverse the elements of $L$
column-wise starting at the upper left corner. Whenever $\alpha$ occurs in
position $(i,j)$ in $L$, then append the number $i + j d$ to the list
$s_{L, \alpha}$. The other ingredient to construct these MUBs is an
arbitrary complex Hadamard matrix $H = (h_{i,j})$ of size $d\times
d$. For each Latin square $L$ and each $\alpha, j \in \{1, \ldots, d\}$
define a normalized vector $\ket{v_{L,\alpha,j}} := 1/\sqrt{d}
\sum_{i=1}^d e_{s_{L,\alpha}[i]} h_{i,j}$, where $e_i$ are the
elementary basis vectors in $\C^{d^2}$. Then the bases given by
$B_L := \{\ket{v_{L,\alpha,j}} : \alpha,j = 1,\ldots,d\}$ together
with the identity matrix $\onemat_{d^2}$ form a set of $w+2$ MUBs.
\smallskip

\begin{example}
In dimension $d=3$ Construction I yields 
the bases
\[
\begin{array}{c@{}rrr}
3^{-1/2} \{& (1,1,1), & (1,\omega_3,\omega_3^2),&  (1,\omega_3^2,\omega_3)\},\\[1ex]
3^{-1/2} \{ &  (1,\omega_3,\omega_3),  
&(1,\omega_3^2,1),  
&(1,1,\omega_3^2)\},\\[1ex]
3^{-1/2} \{ & 
(1,\omega_3^2,\omega_3^2),  
&(1,\omega_3,1),
&(1,1,\omega_3)
\},
\end{array}
\]
which together with the standard basis $\onemat_3$ form a maximal
system of four MUBs in $\C^3$.
\end{example}

\begin{example}
In dimension $d=4$ Construction II yields the bases (where we have
abbreviated $``+"$ for $1$ and $``-"$ for $-1$ and $i = \sqrt{-1}$):
\[ 
\begin{array}{lcl@{}ll}
\frac{1}{2} \{ (+,+,+,+), (+,+,-,-), (+,-,-,+), (+,-,+,-) \},\\[1ex]
\frac{1}{2} \{ (+,-,-i,-i), (+,-,i,i), (+,+,i,-i), (+,+,-i,i) \},\\[1ex]
\frac{1}{2} \{ (+,-i,-i,-), (+,-i,i,+), (+,i,i,-), (+,i,-i,+) \},\\[1ex]
\frac{1}{2} \{ (+,-i,-,-i), (+,-i,+,i), (+,i,+,-i), (+,i,-,i) \}.
\end{array}
\]
These four bases and the standard basis $\onemat_4$ form an extremal
set of five MUBs in $\C^4$.
\end{example}

A basic question is how many bases can be achieved in general
dimension. To this end, we define the function $M: \N \rightarrow \N$
as follows:
\[ 
M(n) := \max\{ |{\cal B}|: \; {\cal B} \; \textup{is a set of
MUBs in} \; \C^n\}
\]
Then we have that:
\begin{itemize}
\item
$M(p^r) = p^r+1$  for $p$ prime, $r \in \N$,
\item
$M(n) \leq n+1$  for all $n\in \N$,
\item
$M(mn) \geq \min\{ M(m), M(n) \}$  for all $m,n \in \N$.
\item
$M(d^2) \geq N(d)$, where $N(d)$ is the number of mutually orthogonal
Latin squares of size $d \times d$.
\end{itemize}

An open problem is to show that $\liminf_{n\rightarrow \infty} M(n) =
\infty$.

\section{Welch's Lower Bounds} 
Suppose that $X$ is a finite nonempty set of vectors of unit norm in
the complex vector space~$\C^d$. The vectors in $X$ satisfy the
inequalities 
\begin{equation}\label{eq:welch}
\frac{1}{|X|^2} \sum_{x,y\in X} |\scal{x}{y}|^{2k} 
\ge \frac{1}{{d+k-1\choose k}},
\end{equation}
for all integers $k\ge 0$. Welch derived these bounds
in~\cite{welch74} to obtain a lower bound on the maximal
cross-correlation of spreading sequences of synchronous code-division
multiple-access systems. Blichfeld~\cite{blichfeld29} and
Sidelnikov~\cite{sidelnikov74} derived similar bounds for real vectors
of unit norm.

A set $X$ attaining the Welch bound (\ref{eq:welch}) for $k=1$ is
called a WBE-sequence set, a notion popularized by Massey and
Mittelholzer~\cite{massey93} and others. Using
equation~(\ref{eq:welch}), it is straightforward to check that the
union of $d+1$ mutually unbiased bases of $\C^d$ form a WBE-sequence
set. These extremal sets of mutually unbiased bases are even better,
since they also attain the Welch bound for $k=2$.  In fact, we show
that a sequence set attains the Welch bounds (\ref{eq:welch}) for all
$k\le t$ if and only if it is a $t$-design in the complex projective
space~$\C P^{d-1}$. 

Let us introduce some notation. 
Let $S^{d-1}$ denote the sphere of unit vectors in the complex vector
space~$\C^d$.  We say that two vectors $u$ and $v$ of $S^{d-1}$ are
equivalent, in signs $u\equiv v$, if and only if $u=e^{i\theta}v$ for
some $\theta \in\mathbf{R}$. It is easy to see that $\equiv$ is an equivalence
relation. We denote the quotient manifold $S^{d-1}\!/\!\equiv$ by $\C
S^{d-1}$.  Notice that the manifold $\C S^{d-1}$ is isomorphic to the
complex projective space $\C P^{d-1}$, but we prefer the former notation 
because normalizing vectors to unit
length is common practice in quantum computing.
\smallskip

\begin{lemma}\label{th:integral} 
Let $\mu$ be the unique normalized $U(d)$-invariant Haar measure on
the complex sphere $\C S^{d-1}$. For any $x\in S^{d-1}$, we have 
$$  \int_{\C S^{d-1}} |\scal{x}{y}|^{2k}d\mu(y)= \frac{1}{{d+k-1\choose k}}.$$
\end{lemma}
\begin{proof}
The unitary group $U(d)$ acts transitively on the manifold $\C S^{d-1}$. 
This means that for any $y\in \C S^{d-1}$ there exists a unitary matrix $U$ 
mapping $y$ to the first basis vector, $Uy=e_1$. Therefore, 
\begin{eqnarray*}
\int_{\C S^{d-1}} |\scal{x}{y}|^{2k} d\mu(x) &=& \int_{\C S^{d-1}}
|\scal{Ux}{e_1}|^{2k} d\mu(x)\\
&=& \int_{\C S^{d-1}} |\scal{x}{e_1}|^{2k}
d\mu(x),
\end{eqnarray*}
where the last equality holds because of the $U(d)$-invariance of the
measure $\mu$. Using Proposition 1.4.9 from Rudin~\cite{rudin80}, we
obtain
$$ 
\int_{\C S^{d-1}} |\scal{x}{e_1}|^{2k} d\mu(x) = 
\int_{\C S^{d-1}} |x_1^k|^2 d\mu(x) = \frac{1}{{d+k-1\choose d-1}},$$
which proves the claim.
\end{proof}

\section{Complex Projective $t$-Designs} 
We now present some background material on complex projective
designs. We will relate those later on to the systems of vectors
formed by a maximal set of MUBs. 

\nix{
We define the acute angle  $0\le \theta_{xy}\le \pi/2$ between
$x,y\in \C P^{d-1}$ by
\[ 
 \cos\theta_{xy} =
\frac{|\scal{x}{y}|}{\sqrt{\scal{x}{x}\scal{y}{y}}}.
\]

Given a set $X$ of vectors in projective space, the corresponding
$A$-set (the set of angles) is defined by $ A = \{ \alpha_{xy} :=
(\cos\theta_{xy})^2 : x,y\in X, x\neq y\}.$ 
}
\nix{
We state the following
theorem from \cite{CS:1998} which we will need for the
characterization of $t$-designs in Theorem \ref{th:wbe} as such sets
in projective space which meet the first $t+1$ of Welch's
inequalities.
 
\begin{theorem}
Let $\mu$ denote the unique $U(d)$-invariant normalized nonzero measure on $\C
S^{d-1}$. Let $X$ be a finite nonempty set. The following conditions
are equivalent:
\begin{itemize}
\item[1)] $\sum_{x\in X} h(x,\overline{x})=0$ for all $h\in \Harm(k,k)$
and all $k$ in the range $1\le k\le t$.
\item[3)] $\displaystyle \sum_{x,y\in X} Q_k(|\scal{x}{y}|^2)=0$ holds for all $k$ in the range $1\le k\le t$. 
\item[4)] $\displaystyle \sum_{y\in X} Q_k(|\scal{x}{y}|^2)=0$ holds for all $x\in X$ and all $k$ in the range $1\le k\le t$.
\item[2)] The following cubature formula holds for all $f$ in $\Hom(t-1,t-1)$ and $\Hom(t,t)$: 
$$\displaystyle \frac{1}{|X|} \sum_{x\in X} f(x) = \int_{\C S^{d-1}} f(x)d\mu(x).$$
\end{itemize}
\end{theorem}
}

Let us first introduce some notation. We denote by $\Hom(k,\ell)$ the
subset of the polynomial ring $\C[x_1,\dots,x_d, y_1,\dots,y_d]$
that consists of all polynomials that are
homogeneous of degree~$k$ in the variables $x_1,\dots,x_d$ and
homogeneous of degree~$\ell$ in the variables $y_1,\dots, y_d$.
We associate to each polynomial $p$ in $\Hom(k,\ell)$ a function
$p_\circ$ on the sphere $S^{d-1}$ by defining
$p_\circ(\xi)=p(\xi,\overline{\xi})$ for $\xi \in S^{d-1}$.  The
function $p_\circ$ is called the ``restriction'' of $p$ onto the complex
sphere.  It follows from the homogeneity conditions of the
polynomial~$p$ that
$p_\circ(e^{i\vartheta}\xi)=e^{i\vartheta(k-\ell)}p_\circ(\xi)$ holds for all 
$\vartheta\in \mathbf{R}$.
Therefore, we obtain a well-defined polynomial function on $\C S^{d-1}$
only if~$k=\ell$. We define $\Hom(k,k)_\circ=\{ p_\circ\colon p\in \Hom(k,k)\}$. 
\smallskip

\begin{definition}
A finite nonempty subset $X$ of $\C S^{d-1}$ is a  $t$-design in
$\C S^{d-1}$  iff the cubature formula
\[ 
 \frac{1}{|X|}\sum_{x\in X} f(x) = \frac{1}{\mu(\C
S^{d-1})}\int_{\C S^{d-1}} f(x) d\mu(x)
\]
holds for all $f$ in $\Hom(t,t)_\circ$. 
\end{definition}
\smallskip

We now show a characterization of $t$-designs in terms of the
inequalities by Welch given in equation~(\ref{eq:welch}).

\smallskip

\begin{theorem}\label{th:wbe}
Suppose that $X$ is a finite nonempty subset of  $\C S^{d-1}$. 
Then the following statements are equivalent: 
\begin{itemize}
\item[1)] The set $X$ is a $t$-design in $\C S^{d-1}$;
\item[2)] for all $x\in \C^d$ and all $k$ in the range $0\le k\le t$, 
we have the equality 
\begin{equation}\label{eq:embedding}
\frac{\scal{x}{x}^k}{{d+k-1\choose k}} = \frac{1}{|X|} \sum_{y\in X}  
|\scal{x}{y}|^{2k} ;
\end{equation}
\item[3)] the set $X$
satisfies the Welch bounds (\ref{eq:welch}) with equality for all $k$ in the
range $0\le k\le t$, that is
\begin{equation}\label{eq:welcheq}
\frac{1}{|X|^2} \sum_{x,y\in X} |\scal{x}{y}|^{2k} = 
\frac{1}{{d+k-1\choose k}}, \qquad 0\le k\le t.
\end{equation}
\end{itemize}
\end{theorem}
\begin{proof}
We show that 1) implies 2). 
Fix a vector $x\in \C^d$. Note that
$p(y)=|\scal{x}{y}|^{2k}=\scal{y}{x}^k\scal{x}{y}^k$ is a polynomial function 
in $\Hom(k,k)_\circ$. Since $X$ is a $t$-design, 
the exact cubature formula
$$ \frac{1}{|X|}\sum_{y\in X} |\scal{x}{y}|^{2k} = \int_{\C S^{d-1}}
|\scal{x}{y}|^{2k} d\mu(y)$$ holds for all $k$ in the range $0\le k\le
t$. 
By Lemma~\ref{th:integral}, the latter integral
evaluates to ${d+k-1 \choose k}^{-1}$, which proves that equation
(\ref{eq:embedding}) holds for all $k\le t$.
\smallskip

We show next that 2) implies 3).  We observe that 
(\ref{eq:embedding}) holds for all $k\le t$, hence summing over $x\in
X$ yields (\ref{eq:welcheq}).
\smallskip

Finally, we show that 3) implies 1). 
Suppose that equation (\ref{eq:welcheq}) holds. For a vector $x\in
\C^d$, we denote by $x^{\otimes k}$ the $k$-fold tensor product
$x^{\otimes k}=x\otimes \cdots \otimes x \in \C^{d^k}$. Note that
$\scal{x^{\otimes k}}{y^{\otimes k}}=\scal{x}{y}^k$. Consider the $d^{2k}$-dimensional vector
$$\xi = \frac{1}{|X|}\sum_{x\in X} 
x^{\otimes k} \otimes \overline{x}^{\otimes k}  - 
\int_{\C S^{d-1}} x^{\otimes k} \otimes \overline{x}^{\otimes k} d\mu(x).
$$  
Evaluating the inner product of $\xi$ with itself yields 
\begin{equation}\label{eq:sidelnikov} 
\frac{1}{|X|^2}\sum_{x,y\in X} |\scal{x}{y}|^{2k}
- \int \int_{\C S^{d-1}} |\scal{x}{y}|^{2k} d\mu(y)d\mu(x),
\end{equation}
which is equal to $\scal{\xi}{\xi} \geq 0$. The inner integral
evaluates to ${d+k-1\choose k}^{-1}$ by Lemma~\ref{th:integral}, and
the double integral has the same value, because the measure $\mu$ is
normalized. It follows from our assumption that the right hand side
vanishes. By construction of $\xi$, we can conclude that averaging
over $X$ yields an exact cubature formula for all monomials in
$\Hom(k,k)_\circ$, hence, by linearity, for all polynomials in
$\Hom(k,k)_\circ$. This means that $X$ is a $t$-design.
\end{proof}
\smallskip

\begin{remark}
Equation (\ref{eq:sidelnikov}) provides a short proof of the Welch
inequalities~(\ref{eq:welch}).  The analogue for real spherical
$t$-designs of the above result is sketched in \cite{goethals81}.  A
connection to the existence of certain isometric Banach space
embeddings is given in~\cite{koenig95}. 
\end{remark}

\section{Uniform Tight Frames} 
A finite subset $F$ of nonzero vectors of $\C^d$ is called a frame if
there exist nonzero real constants $A$ and $B$ such that 
$$ A \|v\|^2 \le \sum_{f\in F} |\scal{f}{v}|^2\le B\|v\|^2$$ holds for
all $v\in \C^d$. The notion of a frame generalizes the concept of an
orthonormal basis. The linear span of the vectors in $F$ the space
$\C^d$, but the vectors in a frame are in general not linearly
independent. A frame is called tight if and only if the frame bounds
$A$ and $B$ are equal. A tight frame is called isometric (or uniform)
if and only if each vector in $F$ has unit norm. 

\begin{theorem}
Let $F$ be a finite nonempty subset of vectors in $\C^d$. 
The following statements about $F$ are equivalent:
\begin{itemize}
\item[1)] $F$ is a uniform tight frame;
\item[2)] $F$ is a WBE-sequence set; 
\item[3)] $F$ is a $1$-design in $\C S^{d-1}$. 
\end{itemize}
\end{theorem}
\begin{proof}
The frame constants of a uniform tight frame $F$ in $\C^d$ are given
by $A=B=|F|/d$, see for example Property 2.3
in~\cite{goyal01}. Therefore, $F$ satisfies
equation~(\ref{eq:embedding}) of Theorem~\ref{th:wbe} for $k=1$. The
equivalence of the three statements follow now from Theorem~\ref{th:wbe}.
\end{proof}

\begin{corollary}
Any $1$-design in $\C P^{d-1}$ is obtained by projecting an orthogonal
basis from a higher-dimensional space (where all basis vectors have the
same norm).
\end{corollary}

\section{Equivalence of MUBs and $2$-Designs}

\nix{
It is known \cite{hoggar84,CS:1998} that a spherical $t$-design in
$X=\C P^{d-1}$ is a finite subset $\mathcal{B}$ of $X$ such that
$$ \sum_{x,y\in \mathcal{B}} Q_k(\cos2\theta_{xy})=0,\qquad
\mbox{for}\quad k=1,2,\dots, t,$$
where $Q_k(x)$ denotes the Jacobi polynomial  
$$ Q_k(x) = \frac{(d)_{2k}}{(k-1)!\cdot k!}\sum_{i=0}^{k} (-1)^i{k \choose i} 
\frac{_i(k)}{_i(2k+d-2)}x^{k-i},$$,
$\cos2\theta_{xy}=|\scal{x}{y}|^2/(\|x\| \|y\|)$ and $(p)_0= {_0\!(p)}=1$, $(p)_a=p(p+1)\cdots (p+a-1)$, 
}

We need a few more notations before we state our main results.  
If $\mathcal{B}$ is a
subset of $\C S^{d-1}$, then the set $A=\{|\langle x|y\rangle|^2 \colon
x,y\in \mathcal{B}, x\neq y\}$ is called the ``angle'' set of
$\mathcal{B}$. For an element $x$ in $\mathcal{B}$ and an ``angle'' $\alpha\in A$, we define the subdegree $d_\alpha(x)$ as 
$d_\alpha(x)=|\{ y \in \mathcal{B}\colon
|\scal{x}{y}|^2=\alpha\}|$.  If the subdegree $d_\alpha$ of an
$\alpha\in A$ is independent of $x$, then $\mathcal{B}$ is called a
\textit{regular scheme}. Note that the union of mutually orthogonal bases of $\C^d$ is a regular scheme with angle set $\{0,1/d\}$. 
\smallskip

\begin{theorem}
The union $X$ of $d+1$ mutually unbiased bases in $\C^d$ forms a
2-design in $\C S^{d-1}$ with angle set $\{0,1/d\}$ and
$d(d+1)$ elements.
\end{theorem}
\smallskip

\begin{proof}
We verify that $X$ attains the Welch bound in equation~(\ref{eq:welch})
with equality for $0\le k\le 2$. The statement then follows from
Theorem \ref{th:wbe}. Indeed, this is obvious for $k=0$. We note that
$|X|=d(d+1)$.

If we evaluate the left hand side of the Welch bound for $X,$ then we obtain 
\begin{eqnarray*}
\frac{1}{d^2(d{+}1)^2} \sum_{x,y\in X} |\langle x|y\rangle|^2 &\!\!\!=\!\!\!& 
\frac{d(d{+}1)}{d^2(d{+}1)^2}\left(1 {+} (d{-}1)0 {+} d^2\frac{1}{d}\right) \\
&=& \frac{1}{d}
\end{eqnarray*}
and this coincides with $\binom{d+1-1}{1}^{-1}=1/d$; so, $X$ is a
1-design. Similarly, for $k=2$,
\begin{eqnarray*}
\frac{1}{d^2(d{+}1)^2} \sum_{x,y\in X} |\langle x|y\rangle|^4\!\! &\!\!\!\!\!\!\!=\!\!\!\!\!\!\!& \!\!
\frac{d(d{+}1)}{d^2(d{+}1)^2}\left(1 {+} (d{-}1)0 {+} d^2\frac{1}{d^2}\right)\\
& =& 
\frac{2}{d(d+1)},
\end{eqnarray*}
and this coincides with $\binom{d+2-1}{2}^{-1}= 2/(d(d+1))$. 
\end{proof}
\smallskip

\begin{theorem}
A 2-design $\mathcal{B}$ in complex projective space $\C
S^{d-1}$ with angle set $\{0,1/d\}$ and $|\mathcal{B}|=d(d+1)$
elements is the union of $d+1$ mutually unbiased bases.
\end{theorem}
\smallskip

\begin{proof}
A complex projective 2-design with $s=|\{0,1/d\}|=2$ satisfies $2\ge
s-1$, hence is a regular scheme \cite{hoggar84}. For $\alpha=1/d$, any
$x\in \mathcal{B}$ has subdegree $d_{\alpha}(x) = d^2$ by Theorem~2.5
of \cite{hoggar84}. It follows that $x$ is orthogonal to $d-1$
elements.

Let $B_x = \{ x\} \cup \{ z \in \mathcal{B}\colon \scal{x}{z}=0\}$.
We claim that $B_x$ is an orthonormal basis of $\C^d$.  We may assume
that the vectors in $\mathcal{B}$ are normalized to unit norm.  Thus,
it suffices to show that $B_x=B_y$ for each $y\in B_x$.  For $x=y$
this is trivial. We know that $x$ and $y$ are contained in both $B_x$
and $B_y$. Therefore, it suffices to show that the intersection set 
$$I(x,y)=\{z\in\mathcal{B}\colon \scal{x}{z}=0, \scal{y}{z}=0\}=B_x\cap B_y -
\{x,y\}$$ contains $d-2$ elements. 

The number of elements in $I(x,y)$ does not depend on $x,y$ for a
$t$-design with $t\ge 2s-2$, see~\cite{hoggar92}. 
Specializing Theorem~5.2 in \cite{hoggar92} to the case at hand shows that 
$$ |I(x,y)| = d^2 \sum_{i,j=0}^1 \sigma_{1-i}^0 \sigma_{1-j}^0
\big(d(d+1) g_{ij}(0)-0^i-0^j\big).$$ We can now evaluate the intersection
polynomials $g_{ij}(0)$ using \cite[Theorem~5.3]{hoggar92} and obtain
that $|I(x,y)|= d-2$.

Hence we can conclude that each set $B_x$ forms an orthonormal basis of
$\C^d$.  The sets $B_x$ partition $\mathcal{B}$. If $B_x\neq B_z$,
then the bases are by construction mutually unbiased. 
\end{proof}

Zauner conjectures that if the dimension $d$ is not a prime power,
then a 2-design with angle set $\{0,1/d\}$ cannot have $d(d+1)$
elements~\cite{Zauner:99}. His conjecture can now be reformulated in terms
of mutually unbiased bases, which then states that $N(d)<d+1$ for
non-prime power $d$.  If Zauner's conjecture is true, then this would
explain the particular role of the finite field construction by
Wootters and Fields~\cite{WF:89}.

\begin{remark} 
Theorem 3 was obtained earlier by Zauner as part of a more general
result on combinatorial quantum designs using a different terminology,
see~\cite[Theorem 2.19]{Zauner:99}. The converse direction, our
Theorem 4, appears to be new. 
\end{remark}

\section{SIC-POVMs and 2-Designs} 

Finally, to demonstrate the versatility of Theorem \ref{th:wbe} we
also show that another system of vectors used in quantum information
theory corresponds to complex projective $2$-designs. So-called
symmetric informationally complete positive operator-valued measures
(SIC-POVMs) are systems of $d^2$ vectors in $\C^d$ which have constant
inner product, i.\,e., $|\langle v, w \rangle|^2 = 1/(d+1)$ for all
$v,w$ in the set. Like in case of MUBs it is a challenging task to
construct SIC-POVMs---indeed here solutions are known only for
a finite number of dimensions \cite{Grassl2004,RBSC:2004}. In
\cite{RBSC:2004} it was shown that SIC-POVMs actually form complex
projective $2$-designs. The following theorem gives a new and simple
proof of this result.
\smallskip

\begin{theorem}[SIC-POVMs are $2$-designs \cite{RBSC:2004}]
Let $X$ be a SIC-POVM $X$ in dimension $d$. Then $X$ forms a
2-design in $\C S^{d-1}$ with angle set $\{1/(d+1)\}$ and
$d^2$ elements.
\end{theorem}
\smallskip

\begin{proof}
Again, we only have to verify that the set $X$ of vectors attains the
Welch bound with equality for $0\le k\le 2$. The statement then
follows from Theorem \ref{th:wbe}. Indeed, this is obvious for
$k=0$. We note that here $|X|=d^2$. Evaluating the left hand side
of the Welch bound for $X,$ then we obtain
\begin{eqnarray*}
\frac{1}{d^4} \sum_{x,y\in X} |\langle x|y\rangle|^2 &\!\!\!=\!\!\!& 
\frac{1}{d^4}\left(d^2 \cdot 1 {+} (d^4-d^2) \frac{1}{d+1}\right) \\
&=& \frac{1}{d^2} \left(1 + (d-1)\right) = \frac{1}{d}
\end{eqnarray*}
and this coincides with $\binom{d+1-1}{1}^{-1}=1/d$; so, $X$ is a
1-design. Similarly, for $k=2$,
\begin{eqnarray*}
\frac{1}{d^4} \sum_{x,y\in X} |\langle x|y\rangle|^4\!\! &\!\!\!\!\!\!\!=\!\!\!\!\!\!\!& \!\!
\frac{1}{d^4}\left(d^2 \cdot 1 + (d^4 - d^2)\frac{1}{(d+1)^2}\right)
\\
& =& 
\frac{1}{d^2}\left( 1 + \frac{d-1}{d+1}\right) = \frac{2}{d(d+1)}
\end{eqnarray*}
and this coincides with $\binom{d+2-1}{2}^{-1}= 2/(d(d+1))$. 
\end{proof}

\begin{remark}
Zauner pointed out to us that the previous theorem can also be
obtained in the language of combinatorial quantum designs by combining
Theorems~2.29 and 2.30 in his dissertation~\cite{Zauner:99}.
\end{remark}

\section{Conclusion}
We have shown that the seemingly unrelated concepts of MUBs on the one
hand and complex projective $2$-deigns on the other are actually the
same objects. This was anticipated in a paper by Barnum
\cite{Barnum:2002} in which it was shown that the union of the (d+1)
bases of a particular system of MUBs forms a complex projective
$2$-design. In the present paper we have generalized this to arbitrary
MUBs and have also shown the other direction, i.\,e., any $2$-design
in dimension $d$ which consists of $d^2+d$ elements and has angle set
$\{0,1/d\}$ can be partitioned into $d+1$ sets of MUBs. We have also
shown that these sets meet the Welsh bounds for $k=0,1,2$ with
equality.  Hence, the present paper can also be seen as a
generalization of the results of \cite{Waldron:2003} in which the
corresponding statement over the real numbers was shown.  Finally, we
would like to mention that Zauner \cite{Zauner:99} conjectures that
affine 2-designs do not exist in dimensions $d$ having two distinct
prime factors.

\section*{Acknowledgment}
We are grateful to Emina Soljanin for bringing tight frames to our
attention, and to Gerhard Zauner for assisting with the translation of
his results. We also thank Chris Godsil, Markus Grassl, Joe Renes,
Aidan Roy for several interesting discussions, and Richard Cleve for 
pointing out a small error in an example of a previous version. 

The research of A.K. was supported in part by NSF grant CCR-0218582,
NSF CAREER award CCF-0347310, a TEES Select Young Faculty award, and a
Texas A\&M TITF grant. This work was carried out while M.R. was at the
Institute for Quantum Computing, University of Waterloo, Canada.

\enlargethispage{-6.7cm} 


\end{document}